\documentclass[submission,copyright,creativecommons]{eptcs}
\providecommand{\event}{F. Ricca, A. Russo et al. (Ed.): Proceedings 36th International Conference\\ 
on Logic Programming (Technical Communications) 2020 (ICLP 2020)\\
EPTCS ??, 2020, pp. 1–14, doi:10.4204/EPTCS.??.??} % Name of the event you are submitting to
\usepackage{breakurl}             % Not needed if you use pdflatex only.
\usepackage{underscore}           % Only needed if you use pdflatex.

\usepackage{mathptmx}
\usepackage{amssymb}
\usepackage[english]{babel}
\usepackage{hyperref}
\usepackage{url}
\usepackage{multirow}
\usepackage{xcolor}
\usepackage{verbatim}
\usepackage{fancyvrb}
\usepackage{xspace}
\usepackage{booktabs}
\usepackage{graphicx}
\usepackage{inconsolata}
\usepackage{float}
\usepackage{amsmath}
\usepackage{breqn}

\newcommand{\fogbrain}{\xspace{\sf\small FogBrain}\xspace}
\newcommand{\reasoner}{\xspace{\sf\small FogBrain}\xspace}
\newcommand{\example}{\noindent\textbf{Example.}\xspace}

\title{Continuous Reasoning for Managing\\ Next-Gen Distributed Applications}
\author{Stefano Forti and Antonio Brogi
\institute{Department of Computer Science, University of Pisa, Italy}
\email{\{stefano.forti, antonio.brogi\}@di.unipi.it}
}
\def\titlerunning{Continuous Reasoning for Managing Next-Gen Distributed Applications}
\def\authorrunning{S. Forti and A. Brogi}

\begin{document}
\maketitle

\begin{abstract}
Continuous reasoning has proven effective in incrementally analysing changes in application codebases within Continuous Integration/Continuous Deployment (CI/CD) software release pipelines.
In this article, we present a novel declarative continuous reasoning approach to support the management of multi-service applications over the Cloud-IoT continuum, in particular when infrastructure variations impede meeting application's hardware, software, IoT or network QoS requirements.
We show how such an approach brings considerable speed-ups compared to non-incremental reasoning.
%  Continuous reasoning has been recently proposed as an effective way to incrementally analyse changes in application codebases within Continuous Integration/Continuous Deployment (CI/CD) software release pipelines.
% %
% In this article, we present a novel declarative continuous reasoning approach to support the runtime management of multi-service applications over the Cloud-IoT continuum, in particular when infrastructure variations impede meeting application's hardware, software, IoT or network QoS requirements.
% %
% We  show how such an approach shows considerable speed-ups with respect to non-incremental reasoning solutions.
\end{abstract}

%\vspace{-6mm}
\section{Introduction}
\label{sec:introduction}

 Large IT companies rely on \textit{continuous reasoning} to support iterative software development up to its \textit{continuous integration} within a single shared codebase~\cite{cr4ci}. For instance, static analyses based on compositional models of programs and code repositories permit exploiting continuous reasoning to check security properties on a whole codebase repository by verifying them only on the changes occurred since the last performed verification. By employing such techniques, Facebook Infer %\footnote{\textit{Infer static analyser}, \url{https://fbinfer.com/}} 
 incrementally and continuously checks whether new commits from Facebook developers are safe (and can be accepted in the codebase) or not (and should be revised before acceptance)~\cite{inferlast}. This dramatically reduces analysis times, instead of repeating a complete codebase analysis at every single commit.

Meanwhile, the Internet of Things (IoT) is undergoing a relentless growth which is becoming more difficult to support with the current software and infrastructure architectures. Besides, many next-gen IoT applications will have QoS requirements such as low latencies, network bandwidth availability and deployment security, which are difficult to guarantee over Cloud-only infrastructures \cite{brogi2019place}. 
Relying upon a large-scale hierarchy of distributed nodes, Cloud-IoT computing paradigms aim at enabling the QoS- and context-aware deployment of next-gen IoT application services to any device supporting them over the Cloud-IoT computing continuum, e.g. in Fog~\cite{bellavistafog} or Osmotic~\cite{DBLP:journals/computer/VillariFDRJR19} computing.
In this scenario, the problem of deciding where to place application services (i.e., functionalities) to infrastructure nodes is of primary importance and it is provably NP-hard~\cite{QoSawaredeployment2017,cardellininphard}. 
In recent years, much literature focussed on determining the best QoS- and context-aware placement of multi-service IoT applications to Cloud-IoT infrastructures, mainly exploiting search-based and mathematical programming solutions~\cite{brogi2019place}.

Inspired from continuous reasoning and from our previous work on next-gen application placement~\cite{fogtorchpibookchapter2019} and management~\cite{fortipagiarobrogi2020}, we consider it promising to extend the concept of continuous reasoning in support of the management of next-gen multi-service distributed applications.
This will reduce the time needed to make management decisions when only part of a running application deployment is affected by changes in the Cloud-IoT infrastructures, e.g. crash of a node hosting a service, degraded network QoS in between communicating services running on different nodes. 
By mainly considering the migration of services suffering due to such infrastructure changes, continuous reasoning can bring two-fold benefits. On one hand, it permits scaling to larger sizes of the placement problem by incrementally solving smaller instances of such a problem, thus reducing the time needed to make informed decisions. On the other hand, it can reduce the number of management operations needed to adapt the current deployment to the new infrastructure conditions, by avoiding unnecessary service migrations. 

In this article, we move a first step towards this direction by introducing the application of continuous reasoning in support of the automated management of next-gen multi-service applications in Cloud-IoT settings. 
We propose a novel declarative methodology and its open-source continuous reasoner prototype, \reasoner, that makes informed service migration decisions, when infrastructure variations prevent running applications from meeting their hardware, software, IoT or network QoS requirements. 
Our methodology shows three main elements of novelty, corresponding to very desirable properties for Cloud-IoT application management support:
\begin{itemize}
    \item it is \textit{declarative}, hence more concise, easier to understand, modify and maintain when compared to existing procedural solutions, and it is also characterised by a high level of flexibility and extensibility, which suits the ever-changing needs of Cloud-IoT scenarios,
    \item it is intrinsically \textit{explainable} as it derives proofs for input user queries by relying on Prolog state-of-the-art resolution engines, and it can be easily extended to justify \textit{why} a certain management decision was taken at runtime in the spirit of explainable AI (XAI), and
    \item it features \textit{scalability} by supporting application management at the large-scale, by relying on a continuous reasoning approach to reduce the size of the considered problem instance only to those application services currently in need for attention.
\end{itemize}

\noindent
The rest of this paper is organised as follows. We first illustrate how \fogbrain determines context- and QoS-aware placements of multiservice applications to Cloud-IoT infrastructures (Sect.~\ref{sec:placement}), and how it implements continuous reasoning for runtime placement decisions (Sect.~\ref{sec:continuousreasoning}). We then discuss the scalability of \fogbrain over examples of increasing size (Sect.~\ref{sec:experiments}), and survey some related work (Sect.~\ref{sec:related}). We finally conclude and highlight directions for future work (Sect.~\ref{sec:conclusions}).

\vspace{-3mm}
\section{Placement of Next-Gen Distributed Applications}
\label{sec:placement}

Hereafter, we describe a declarative solution to the problem of placing multi-service applications to Cloud-IoT computing infrastructures in a context- and QoS-aware manner.
Eligible 
%solutions to such a problem are 
 placements are mappings of each service of an application to a computational node in the available infrastructure, which supports all service software, hardware, IoT and QoS requirements. Placements can include nodes that host more than one service, if their capabilities are sufficient.
The Prolog program described in this section will be the core of the \reasoner prototype that we will present in Sect. \ref{sec:continuousreasoning}.

\smallskip\noindent\textit{Declaring Application Requirements. } First, applications identified by \texttt{\small  AppId} and made from services \texttt{\small  ServiceId1, ..., ServiceIdK} are declared as in

\vspace{-1mm}
\begin{Verbatim}[fontfamily=zi4, fontsize=\footnotesize, frame=single, framesep=1mm, framerule=0.1pt, rulecolor=\color{gray}]
app(AppId, [ServiceId1, ..., ServiceIdK]).
\end{Verbatim}

\vspace{-2mm}
\noindent Second, application services identified by \texttt{\small ServiceId} and associated with their software \texttt{\small SwReqs}, hardware \texttt{\small HwReqs} and IoT requirements \texttt{\small TReqs} are declared as in

\vspace{-1mm}
\begin{Verbatim}[fontfamily=zi4, fontsize=\footnotesize, frame=single, framesep=1mm, framerule=0.1pt, rulecolor=\color{gray}]
service(ServiceId, SwReqs, HwReqs, TReqs).
\end{Verbatim}

\vspace{-3mm}
\noindent Finally, interactions between services \texttt{\small ServiceId1} and \texttt{\small ServiceId2}, associated with their maximum latency \texttt{\small LatReq} and minimum bandwidth \texttt{\small BwReq} requirements are declared as in

\vspace{-1mm}
\begin{Verbatim}[fontfamily=zi4, fontsize=\footnotesize, frame=single, framesep=1mm, framerule=0.1pt, rulecolor=\color{gray}]
s2s(ServiceId1, ServiceId2, LatReq, BwReq).
\end{Verbatim}

\begin{figure}[H]
    \centering
    \includegraphics[width=0.9\textwidth]{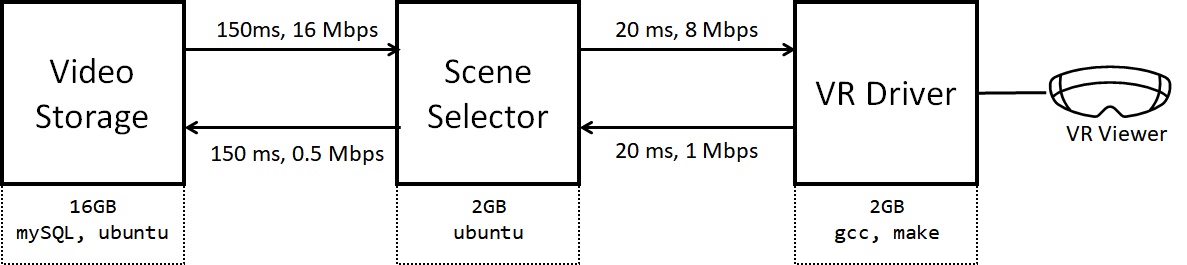}
    \caption{VR application of the example.}
    \label{fig:vrapp}
\end{figure}

\example Consider the Virtual Reality (VR) application in Fig. \ref{fig:vrapp}, which is made of three interacting services. VR videos are streamed from the {\sf\small Video Storage} service to the {\sf\small VR Driver} service, connected to a VR Viewer device, passing through the {\sf\small Scene Selector} service, which selects the portion of the video to show to the user, based on her currently sensed head positioning.

\noindent The requirements of such an application can be easily declared\footnote{For the sake of simplicity, we consider generic hardware units to express the hardware requirements to be met, as in most of the approaches we surveyed in \cite{brogi2019place}.} as in Fig. \ref{fig:vrappcode}.

\begin{figure}[H]
    \centering
\begin{Verbatim}[fontfamily=zi4, fontsize=\footnotesize, frame=single, framesep=1mm, framerule=0.1pt, rulecolor=\color{gray}]
application(vrApp, [videoStorage, sceneSelector, vrDriver]).
service(videoStorage, [mySQL, ubuntu], 16, []).
service(sceneSelector, [ubuntu], 2, []).
service(vrDriver, [gcc, make], 2, [vrViewer]).
s2s(videoStorage, sceneSelector, 150, 16).
s2s(sceneSelector, videoStorage, 150, 0.5).
s2s(sceneSelector, vrDriver, 20, 8).
s2s(vrDriver, sceneSelector, 20, 1).
\end{Verbatim}
    \caption{Example VR application specification.}
    \label{fig:vrappcode}
\end{figure}

\noindent
For instance, the {\sf\small VR Driver} service requires both \texttt{\small gcc} and \texttt{\small make} to bootstrap on the target deployment node, the availability of $2$ hardware units and the reachability of the specified {\sf\small vrViewer}. Analogously, the service-to-service link between the {\sf\small Scene Selector} and the {\sf\small VR Driver} requires end-to-end download bandwidth of at least 8 Mbps with at most 20 ms latency to stream videos to the VR Viewer, and end-to-end upload bandwidth of at least 1 Mbps with at most 20 ms latency to receive the currently sensed user's head positioning (as per the last two \texttt{\small s2s/4} facts in Fig. \ref{fig:vrappcode}).
$\hfill \diamond$

\smallskip\noindent\textit{Declaring Infrastructure Capabilities. } Dually to application requirements, it is possible to declare an infrastructure node identified by \texttt{\small NodeId} and its software (i.e. \texttt{\small SwCaps}), hardware (i.e. \texttt{\small HwCaps}) and IoT (i.e. \texttt{\small TCaps}) capabilities, obtained via infrastructure monitoring, as in
\begin{Verbatim}[fontfamily=zi4, fontsize=\footnotesize, frame=single, framesep=1mm, framerule=0.1pt, rulecolor=\color{gray}]
node(NodeId, SwCaps, HwCaps, TCaps).
\end{Verbatim}
\noindent Besides, it is possible to declare the monitored end-to-end latency (i.e. \texttt{\small FeatLat}) and bandwidth (i.e. \texttt{\small FeatBw}) featured by communication links between a pair of nodes \texttt{\small NodeIdA}--\texttt{\small NodeIdB} as in
\begin{Verbatim}[fontfamily=zi4, fontsize=\footnotesize, frame=single, framesep=1mm, framerule=0.1pt, rulecolor=\color{gray}]
link(NodeIdA, NodeIdB, FeatLat, FeatBw).
\end{Verbatim}

\begin{figure}[H]
    \centering
    \includegraphics[width=0.95\textwidth]{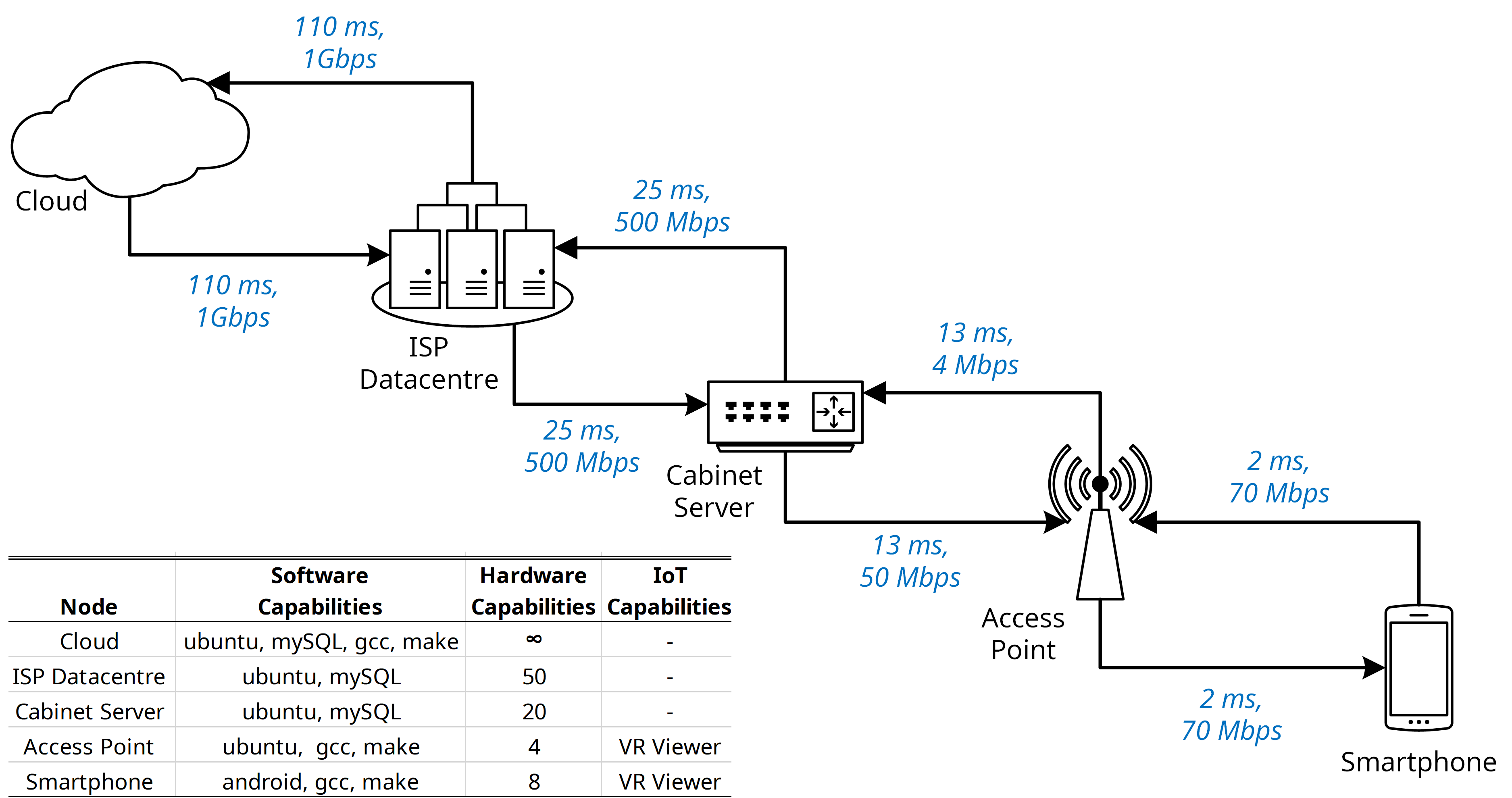}
    \caption{Infrastructure of the example.}
    \label{fig:vrinfra}
\end{figure}

\example Consider the infrastructure of Fig. \ref{fig:vrinfra}, sketching a Cloud-IoT computing continuum that spans through the Internet backbone, an ISP infrastructure, a metropolitan access network, and eventually reaches a home access point enabling a wireless LAN. Assume that both the access point and the smartphone can reach out a VR Viewer in the wireless LAN enabled by the access point. Such VR Viewer can be exploited by a running instance of the application of Fig. \ref{fig:vrapp}.

\vspace{-3mm}

\vspace{-3mm}
\noindent
Such an infrastructure can be represented, including end-to-end links not shown in Fig. \ref{fig:vrinfra}, as listed in Fig. \ref{fig:vrinfracode}.

\begin{figure}[H]
    \centering
\begin{Verbatim}[fontfamily=zi4, fontsize=\footnotesize, frame=single, framesep=1mm, framerule=0.1pt, rulecolor=\color{gray}]
node(cloud, [ubuntu, mySQL, gcc, make], inf, []).
node(ispdatacentre, [ubuntu, mySQL], 50, []).
node(cabinetserver, [ubuntu, mySQL], 20, []).
node(accesspoint, [ubuntu, gcc, make], 4, [vrViewer]).
node(smartphone, [android, gcc, make], 8, [vrViewer]).

link(cloud, ispdatacentre, 110, 1000).            link(cabinetserver, accesspoint, 13, 50).
link(cloud, cabinetserver, 135, 100).             link(cabinetserver, smartphone, 15, 35).
link(cloud, accesspoint, 148, 20).                link(accesspoint, cloud, 148, 3).
link(cloud, smartphone, 150, 18 ).                link(accesspoint, ispdatacentre, 38, 4).
link(ispdatacentre, cloud, 110, 1000).            link(accesspoint, cabinetserver, 13, 4).
link(ispdatacentre, cabinetserver, 25, 500).      link(accesspoint, smartphone, 2, 70).
link(ispdatacentre, accesspoint, 38, 50).         link(smartphone, cloud, 150, 2).
link(ispdatacentre, smartphone, 40, 35).          link(smartphone, ispdatacentre, 40, 2.5).
link(cabinetserver, cloud, 135, 100).             link(smartphone, cabinetserver, 15, 3).
link(cabinetserver, ispdatacentre, 25, 500).      link(smartphone, accesspoint, 2, 70).
\end{Verbatim}

    \caption{Example infrastructure declaration.}
    \label{fig:vrinfracode}
\end{figure}

\noindent
The proposed representation captures asymmetric links (i.e. with different upload and download bandwidth), which are common in Cloud-IoT scenarios, especially in edge networks.
$\hfill \diamond$
 
\smallskip\noindent
\textit{Determining Eligibile Application Placements. } It is now possible to define a declarative \textit{generate \& test} strategy to match application requirements to infrastructure capabilities and determine eligible placements as per the code of Fig. \ref{fig:placement}.
The \texttt{\small placement/2} predicate (lines 1--4) retrieves an application \texttt{\small App} (line 2) and looks for a \texttt{\small Placement} of its \texttt{\small Services} that meets all the application requirements over the available infrastructure by calling the predicate \texttt{\small placement/7} (line 3). A found eligible placement is asserted as a fact into the knowledge base of the \fogbrain reasoner along with the associated hardware \texttt{\small AllocHW} and bandwidth \texttt{\small AllocBW} allocations (line 4).

\begin{figure}[!p]
    \centering
\begin{Verbatim}[fontfamily=zi4, numbers=left, numbersep=5pt, fontsize=\footnotesize, numberblanklines=false, 
frame=single, 
framesep=1mm, framerule=0.1pt, rulecolor=\color{gray},  firstnumber=1,tabsize=2]
placement(App, Placement) :-
	application(App, Services),
	placement(Services, [], AllocHW, [], AllocBW, [], Placement),
	assert(deployment(App, Placement, AllocHW, AllocBW)).

placement([], AllocHW, AllocHW, AllocBW, AllocBW, Placement, Placement).
placement([S|Ss], AllocHW, NewAllocHW, AllocBW, NewAllocBW, Placement, NewPlacement) :-
	servicePlacement(S, AllocHW, TAllocHW, N),
	flowOK(S, N, Placement, AllocBW, TAllocBW), 
	placement(Ss, TAllocHW, NewAllocHW, TAllocBW, NewAllocBW, [on(S,N)|Placement], NewPlacement).

servicePlacement(S, AllocHW, NewAllocHW, N) :-
	service(S, SWReqs, HWReqs, TReqs),
	node(N, SWCaps, HWCaps, TCaps),
	hwTh(T), HWCaps >= HWReqs + T,
	thingReqsOK(TReqs, TCaps),
	swReqsOK(SWReqs, SWCaps),
	hwReqsOK(HWReqs, HWCaps, N, AllocHW, NewAllocHW).

thingReqsOK(TReqs, TCaps) :- subset(TReqs, TCaps).

swReqsOK(SWReqs, SWCaps) :- subset(SWReqs, SWCaps).

hwReqsOK(HWReqs, HWCaps, N, [], [(N,HWReqs)]) :-
	hwTh(T), HWCaps >= HWReqs + T.
hwReqsOK(HWReqs, HWCaps, N, [(N,AllocHW)|L], [(N,NewAllocHW)|L]) :-
	NewAllocHW is AllocHW + HWReqs, hwTh(T), HWCaps >= NewAllocHW + T.
hwReqsOK(HWReqs, HWCaps, N, [(N1,AllocHW)|L], [(N1,AllocHW)|NewL]) :-
	N \== N1, hwReqsOK(HWReqs, HWCaps, N, L, NewL).

flowOK(S, N, Placement, AllocBW, NewAllocBW) :-
	findall(n2n(N1,N2,ReqLat,ReqBW), interested(N1,N2,ReqLat,ReqBW,S,N,Placement), Ss),
	serviceFlowOK(Ss, AllocBW, NewAllocBW).

interested(N, N2, ReqLat, ReqBW, S, N, Placement) :-
	s2s(S, S2, ReqLat, ReqBW), member(on(S2,N2), Placement), N\==N2.
interested(N1, N, ReqLat, ReqBW, S, N, Placement) :-
	s2s(S1, S, ReqLat, ReqBW), member(on(S1,N1), Placement), N\==N1.

serviceFlowOK([], AllocBW, AllocBW).
serviceFlowOK([n2n(N1,N2,ReqLat,ReqBW)|Ss], AllocBW, NewAllocBW) :-
	link(N1, N2, FeatLat, FeatBW),
	FeatLat =< ReqLat,
	bwOK(N1, N2, ReqBW, FeatBW, AllocBW, TAllocBW),
	serviceFlowOK(Ss, TAllocBW, NewAllocBW).

bwOK(N1, N2, ReqBW, FeatBW, [], [(N1,N2,ReqBW)]):-
	bwTh(T), FeatBW >= ReqBW + T.
bwOK(N1, N2, ReqBW, FeatBW, [(N1,N2,AllocBW)|L], [(N1,N2,NewAllocBW)|L]):-
	NewAllocBW is ReqBW + AllocBW, bwTh(T), FeatBW >= NewAllocBW + T.
bwOK(N1, N2, ReqBW, FeatBW, [(N3,N4,AllocBW)|L], [(N3,N4,AllocBW)|NewL]):-
	\+ (N1 == N3, N2 == N4), bwOK(N1,N2,ReqBW,FeatBW,L,NewL).
\end{Verbatim}
    \caption{Multi-service application placement to infrastructure nodes.}
    \label{fig:placement}
\end{figure}

The \texttt{\small placement/7} predicate (lines 5--9) recursively scans the list of services to be deployed (i.e. \texttt{\small [S|Ss]}) to determine an eligible \texttt{\small Placement} for each of them. The variables \texttt{\small AllocHW}/\texttt{\small NewAllocHW} and \texttt{\small AllocBW}/\texttt{\small NewAllocBw} are used to keep track of the hardware and bandwidth allocated by the placement being built in \texttt{\small Placement}/\texttt{\small NewPlacement} (line 6). This information is used not to exceed the hardware capabilities nor to saturate communication links with cumulative hardware or bandwidth requirements of application services mapped onto a same node or link, respectively. When a valid placement for service \texttt{\small S} on node \texttt{\small N} is found, it is appended to the placement being built as the tuple \texttt{\small on(S,N)} (line 9).
To this end, \texttt{\small placement/7} coordinates the checks on service requirements (line 7) with the checks on service-to-service communication requirements (line 8) in two repeated steps, until the list of services to be deployed is empty (line 5).

As a first step, \texttt{\small servicePlacement/4} (lines 7, 10--16) non-deterministically places service \texttt{\small S} onto a node \texttt{\small N} (lines 7, 10--16), and checks that node \texttt{\small N} can at least support the hardware requirements of \texttt{\small S} (line 13) and meets its IoT (lines 14, 17) and software (lines 15, 18) requirements. Then, by means of \texttt{\small hwReqsOK/5} (lines 16, 19--24), \texttt{\small servicePlacement/4} performs a check on cumulative hardware allocation, as it updates the \texttt{\small AllocHW} accumulator into the \texttt{\small NewAllocHW} accumulator by summing up the requirements of \texttt{\small S} to the hardware previously allocated to other services mapped onto \texttt{\small N} as per \texttt{\small Placement}. Thus, the hardware \texttt{\small AllocHW} currently allocated at any node \texttt{\small N} by the current placement can always be found within the list \texttt{\small TAllocHW} (line 7) in the couple \texttt{\small (N, AllocHW)}.

As a second step, \texttt{\small flowOK/5} (lines 8, 25--27) checks whether service-to-service interactions between the last-placed service \texttt{\small S} and previously placed services in \texttt{\small Placement} can support the required network QoS. 
All the requirements of latency and bandwidth to be checked on the communication link between the nodes \texttt{\small N} and \texttt{\small N2} (or \texttt{\small N1} and \texttt{\small N}) that support communicating services \texttt{\small S} and \texttt{\small S2} (or \texttt{\small S1} and \texttt{\small S}), respectively, are retrieved by exploiting the \texttt{\small interested/8} predicate (lines 26, 28--31). 
Then, the \texttt{\small serviceFlowOK/3} (line 27, 32--37) recursively scans the \texttt{\small N2Ns} list, and it checks latency requirements (line 35) and cumulative bandwidth allocation (via \texttt{\small bwOK/6}, line 36, 38--43), and updates the \texttt{\small AllocBw} accumulator accordingly. Overall, this second step implements a \textit{fail-fast} heuristic, by immediately backtracking from a placement that cannot meet requirements on communication QoS with other services.

Finally, it is worth noting that \texttt{\small hwReqsOK/5} and \texttt{\small bwOK/6} rely upon two user-set threshold facts
\begin{Verbatim}[fontfamily=zi4, fontsize=\footnotesize, frame=single, framesep=1mm, framerule=0.1pt, rulecolor=\color{gray}]
hwTh(THW). % 0.5 by default
bwTh(TBW). % 0.2 by default
\end{Verbatim}
\vspace{-1mm}
\noindent which represent the amount of hardware and bandwidth not to be allocated so to avoid overloading nodes and links, respectively.

\medskip
\example By querying the \texttt{\small placement/2} predicate on the example application (Fig. \ref{fig:vrapp}) and infrastructure (Fig. \ref{fig:vrinfra}), it outputs 
\begin{Verbatim}[fontfamily=zi4, fontsize=\footnotesize, frame=single, framesep=1mm, framerule=0.1pt, rulecolor=\color{gray}]
1 ?- placement(vrApp, P).
P = [on(vrDriver, accesspoint), on(sceneSelector, cabinetserver), on(videoStorage, cloud)] 
\end{Verbatim}
\noindent
The obtained placement is one out of the $5^3=125$ that are combinatorially possible, and one of the $12$ eligible ones. DevOps engineers managing the VR application can select this placement to initially deploy it to the available infrastructure. $\hfill \diamond$

\smallskip
\noindent
We conclude this section with a note on the time complexity of the proposed solution. Having to place $S$ services onto $N$ nodes, our prototype could explore at most the whole search space to determine an eligible placement (if any), incurring in a worst-case time complexity of $O(N^S)$. %Such a complexity, due to the combinatorial nature of the problem, can be tamed by exploiting suitable sorting of nodes, e.g. from more capable to less capable nodes, and of application services, e.g. from more resource-demanding to less resource-demanding, as in} \cite{QoSawaredeployment2017}.

\vspace{-5mm}
\section{Continuous Management of Next-Gen Distributed Applications}
\label{sec:continuousreasoning}

In this section, we illustrate how \fogbrain implements continuous reasoning on application placement decisions. We only describe the main predicates of \fogbrain realising such a feature. The complete \fogbrain codebase and interactive documentation, realised with Klipse \cite{klipse}, is available online\footnote{{\scriptsize\sf FogBrain} is available at \url{https://pages.di.unipi.it/forti/fogbrain}}.

%\smallskip
\fogbrain works as per the high-level behaviour of \texttt{\small fogBrain/2} shown in Fig. \ref{fig:fogbrainov}. We assume that \fogbrain is embedded in an infinite loop and that changes in the infrastructure conditions trigger a query to \texttt{\small fogBrain(App, Placement)}, which distinguishes the case of runtime application management (lines 44--46) from the case of first application deployment (lines 47--49).
In the former case, since a \texttt{\small deployment/4} has already been asserted for \texttt{\small App} (line 45), \fogbrain triggers a continuous reasoning step on the existing deployment so to understand which management actions are possibly needed, if the deployment does not currently meet all application requirements (line 46). 
In the latter case, where no previous deployment of \texttt{\small App} exists (line 48), \fogbrain queries the \texttt{\small placement/2} predicate of Fig. \ref{fig:placement} (line 49) to determine a first placement for the input application.

\begin{figure}[!ht]
    \centering
\begin{Verbatim}[firstnumber=44,fontfamily=zi4, numbers=left, numbersep=5pt, fontsize=\footnotesize, frame=single, framesep=1mm, framerule=0.1pt, rulecolor=\color{gray}, tabsize=4]
fogBrain(App, NewPlacement) :-
	deployment(App, Placement, AllocHW, AllocBW),
	reasoningStep(App, Placement, AllocHW, AllocBW, NewPlacement).
fogBrain(App, Placement) :-
	\+ deployment(App,_,_,_),
	placement(App, Placement).
\end{Verbatim}
 \caption{Overview of \fogbrain.}
    \label{fig:fogbrainov}
    \end{figure} 

\noindent The key idea of \fogbrain is to reduce the time needed to make informed application management decisions by limiting the size of the considered placement problem instance, so to tame its NP-hard nature and worst-case exp-time complexity. 
This is achieved by building suitable partially ground queries for \texttt{\small placement/7} via the \texttt{\small reasoningStep/5} predicate, which sets to unbound the placement of those application services that, due to the last infrastructure changes, cannot meet their hardware, software or QoS requirements anymore.
Then, a new placement will be determined only for those application services suffering from the current node or network conditions. As a consequence, a problem instance to be solved is handled much faster in most of the cases. Only when it will not be possible to determine an eligible migration for the suffering services, \fogbrain will look for a new complete placement.

 We now detail how \texttt{\small reasoningStep/5}, shown in Fig. \ref{fig:problemfinder}, works.
 The \texttt{\small reasoningStep/5} predicate (lines 50--52) determines via \texttt{\small toMigrate/2} (lines 51, 53--56) which \texttt{\small ServicesToMigrate} in the current \texttt{\small Placement} need to be migrated due to node or network conditions that impede meeting application requirements. 
Then, it exploits the \texttt{\small replacement/7} (line 52) to query the \texttt{\small placement/7} with partially ground placement information, so to attempt migrating only suffering services. %We will now detail how this two-step reasoning works. 

\begin{figure}[!ht]
    \centering
\begin{Verbatim}[fontfamily=zi4, numbers=left, numbersep=5pt, fontsize=\footnotesize,numberblanklines=false,  frame=single, framesep=1mm, framerule=0.1pt, rulecolor=\color{gray},numberblanklines=false, firstnumber=50, tabsize=2]
reasoningStep(App, Placement, AllocHW, AllocBW, NewPlacement) :-
    toMigrate(Placement, ServicesToMigrate),
    replacement(App, ServicesToMigrate, Placement, AllocHW, AllocBW, NewPlacement).

toMigrate(Placement, ServicesToMigrate) :-
    findall((S,N,HWReqs), onSufferingNode(S,N,HWReqs,Placement), ServiceDescr1),
    findall((SD1,SD2), onSufferingLink(SD1,SD2,Placement), ServiceDescr2),
    merge(ServiceDescr2, ServiceDescr1, ServicesToMigrate).

onSufferingNode(S, N, HWReqs, Placement) :-  
    member(on(S,N), Placement),
    service(S, SWReqs, HWReqs, TReqs),
    nodeProblem(N, SWReqs, TReqs).

onSufferingLink((S1,N1,HWReqs1),(S2,N2,HWReqs2),Placement) :-
    member(on(S1,N1), Placement), member(on(S2,N2), Placement), N1 \== N2,
    s2s(S1, S2, ReqLat, _),
    communicationProblem(N1, N2, ReqLat),
    service(S1, _, HWReqs1, _),
    service(S2, _, HWReqs2, _).
\end{Verbatim}
 \caption{The \texttt{\small reasoningStep/5} predicate.}
    \label{fig:problemfinder}
    \end{figure}   

The \texttt{\small toMigrate/2} predicate (lines 53--56) retrieves all service descriptors of those services that are currently suffering due to node or communication issues via predicates \texttt{\small onSufferingNode/4} (line 54) and \texttt{\small onSufferingLink/3} (line 55), respectively. Service descriptors are triples \texttt{\small (S,N,HWReqs)} containing the service identifier, the current deployment node and the service hardware requirements.

First, the \texttt{\small toMigrate/2} predicate calls \texttt{\small onSufferingNode/4} (lines 57--60) to determine all services suffering due to node changes. In turn, \texttt{\small onSufferingNode/4} exploits \texttt{\small nodeProblem/3} (line 60) to check whether the node onto which a service \texttt{\small S} is deployed as per the current \texttt{\small Placement} (lines 58--59) cannot satisfy all requirements of \texttt{\small S}. 
Analogously, to determine all pairs of service descriptors \texttt{\small (SD1, SD2)} suffering from communication problems, \texttt{\small toMigrate/2} calls \texttt{\small onSufferingLink/3} (lines 61-66). In turn, \texttt{\small onSufferingLink/3} exploits \texttt{\small communicationProblem/3} (line 69) to check if the link supporting the communication between \texttt{\small S1} and \texttt{\small S2} (lines 62--63) does not feature suitable QoS. %features too high latency, is currently overloaded (lines 72--74), or is not available anymore due to a failure (lines 75--76).

Both \texttt{\small nodeProblem/3} and \texttt{\small communicationProblem/3} can be flexibly defined by \fogbrain users to check different properties. Fig. \ref{fig:problems} shows the two default definitions of the predicates checking node and links current capabilities against software, IoT and communication QoS requirements of the application services as well as nodes and links overloading situations (lines 67--69 and 72--73). The default definitions also check nodes and links for failures (lines 70--71 and 75--76, respectively).

\begin{figure}[!ht]
    \centering
\begin{Verbatim}[fontfamily=zi4, numbers=left, numbersep=5pt, fontsize=\footnotesize,numberblanklines=false,  frame=single, framesep=1mm, framerule=0.1pt, rulecolor=\color{gray},numberblanklines=false, firstnumber=67, tabsize=2]
nodeProblem(N, SWReqs, TReqs) :-
    node(N, SWCaps, HWCaps, TCaps),
    hwTh(T), \+ (HWCaps > T, thingReqsOK(TReqs,TCaps), swReqsOK(SWReqs,SWCaps)).
nodeProblem(N, _, _) :- 
    \+ node(N, _, _, _).
    
communicationProblem(N1, N2, ReqLat) :- 
    link(N1, N2, FeatLat, FeatBW), 
    (FeatLat > ReqLat; bwTh(T), FeatBW < T).
communicationProblem(N1,N2,_) :- 
    \+ link(N1, N2, _, _).
\end{Verbatim}
 \caption{Default  \texttt{\small nodeProblem/3} and \texttt{\small communicationProblem/3} predicates.}
    \label{fig:problems}
    \end{figure} 
    
The \texttt{\small merge/3} predicate (line 56) merges into the list \texttt{\small ServicesToMigrate} all found descriptors of suffering services, by removing duplicates and splitting descriptor couples determined by \texttt{\small onSufferingLink/3}.
Such a list, along with the current \texttt{\small Placement}, hardware and bandwidth allocations, is passed to predicate \texttt{\small replacement/6} (line 52), with the goal of determining a \texttt{\small NewPlacement} for those suffering services. 

We now describe the functioning of \texttt{\small replacement/6}, listed in Fig. \ref{fig:querybuilding}.
In case no service needs to be migrated, the \texttt{\small NewPlacement} does not change, i.e. it coincides with the current \texttt{\small Placement} (line 77).
Otherwise, the current \texttt{\small deployment} is retracted, and \fogbrain builds a suitable partially-ground query to \texttt{\small placement/7} (line 84) by means of the predicates \texttt{\small partialPlacement/3} (lines 80--81), \texttt{\small freeHWAllocation/3} (lines 82) and \texttt{\small freeBWAllocation/4} (lines 83).

\begin{figure}[!ht]
    \centering
\begin{Verbatim}[fontfamily=zi4, numbers=left, numbersep=5pt, fontsize=\footnotesize, numberblanklines=false, firstnumber=77, frame=single, framesep=1mm, framerule=0.1pt, rulecolor=\color{gray}, tabsize=2]
replacement(_, [], Placement, _, _, Placement).
replacement(A, ServicesToMigrate, Placement, AllocHW, AllocBW, NewPlacement) :-
    ServicesToMigrate \== [], retract(deployment(A, Placement, _, _)),
    findall(S, member((S,_,_), ServicesToMigrate), Services),
    partialPlacement(Placement, Services, PPlacement),
    freeHWAllocation(AllocHW, PAllocHW, ServicesToMigrate),
    freeBWAllocation(AllocBW, PAllocBW, ServicesToMigrate, Placement), 
    placement(Services, PAllocHW, NewAllocHW, PAllocBW, NewAllocBW, PPlacement, NewPlacement),
    assert(deployment(A, NewPlacement, NewAllocHW, NewAllocBW)).

partialPlacement([],_,[]).
partialPlacement([on(S,_)|P],Services,PPlacement) :-
    member(S,Services), partialPlacement(P,Services,PPlacement).
partialPlacement([on(S,N)|P],Services,[on(S,N)|PPlacement]) :-
    \+member(S,Services), partialPlacement(P,Services,PPlacement).
\end{Verbatim}
 \caption{The \texttt{\small replacement/7} and \texttt{\small partialPlacement/3} predicates.}
    \label{fig:querybuilding}
    \end{figure}

\noindent First of all, \texttt{\small partialPlacement/3} (lines 86--90) recursively scans the current \texttt{\small Placement} and removes from it all services to be migrated so to build the partial placement \texttt{\small PPlacement}.
Then, the predicate \texttt{\small freeHWAllocation/3}, listed in Fig. \ref{fig:hwcomputation}, cleans up the current hardware allocation by removing the hardware allocation of suffering services from the nodes whose services are migrated. This is done by recursively scanning the current allocation for each node \texttt{\small (N, AllocHW)}, summing all hardware requirements of the descriptors of the services to be migrated from \texttt{\small N} via \texttt{\small sumNodeHWToFree} (line 93, 97--99), and by removing them from \texttt{\small AllocHW} (line 94--95). All updated hardware allocations are assembled together into \texttt{\small NewAllocHW}  via \texttt{\small assemble/3} (lines 100--101), while removing zero-hardware allocations (line 100).
Analogously, the \texttt{\small freeBWAllocation/4} cleans up the current bandwidth allocation \texttt{BWAlloc} by removing from communication links all bandwidth allocations that involve at least one suffering service. 

\begin{figure}[!ht]
    \centering
\begin{Verbatim}[fontfamily=zi4, numbers=left, numbersep=5pt, fontsize=\footnotesize, numberblanklines=false, firstnumber=91, frame=single, framesep=1mm, framerule=0.1pt, rulecolor=\color{gray}, tabsize=2]
freeHWAllocation([], [], _).
freeHWAllocation([(N,AllocHW)|L], NewL, ServicesToMigrate) :-
    sumNodeHWToFree(N, ServicesToMigrate, HWToFree),
    NewAllocHW is AllocHW - HWToFree, 
    freeHWAllocation(L, TempL, ServicesToMigrate),
    assemble((N,NewAllocHW), TempL, NewL).

sumNodeHWToFree(_, [], 0).
sumNodeHWToFree(N, [(_,N,H)|STMs], Tot) :- sumNodeHWToFree(N, STMs, HH), Tot is H+HH.
sumNodeHWToFree(N, [(_,N1,_)|STMs], H) :- N \== N1, sumNodeHWToFree(N, STMs, H).
 
assemble((_,NewAllocHW), L, L) :- NewAllocHW=:=0.
assemble((N, NewAllocHW), L, [(N,NewAllocHW)|L]) :- NewAllocHW>0.
\end{Verbatim}
 \caption{The \texttt{\small freeHWAllocation/3} predicate.}
    \label{fig:hwcomputation}
    \end{figure}  

Finally, as listed in Fig. \ref{fig:querybuilding}, the partial placement \texttt{\small PPlacement}, the partial hardware and bandwidth allocations, \texttt{\small PAllocHW} and \texttt{\small PAllocBW}, and the identifiers of the \texttt{\small Services} to be migrated are input to \texttt{\small placement/7} to determine an eligible \texttt{\small NewPlacement} (and the related \texttt{\small NewAllocHW} and \texttt{\small NewAllocBW}) (line 84) to be asserted as a \texttt{\small deployment/4} for the application \texttt{\small A} (line 85).
If partially migrating application services fails, no \texttt{\small deployment/4} is asserted and a complete new placement is looked for as per the second clause of \texttt{\small fogBrain/2} (line 47--49, Fig. \ref{fig:fogbrainov}).

\smallskip
\example Retaking the example of Sect. \ref{sec:placement}, we now exploit \fogbrain to determine a first eligible deployment for the VR application:
%
%\vspace{-0.7mm}
\begin{Verbatim}[fontfamily=zi4, fontsize=\footnotesize, frame=single, framesep=1mm, framerule=0.1pt, rulecolor=\color{gray}]
1 ?- fogBrain(vrApp,P).        
P = [on(vrDriver, accesspoint), on(sceneSelector, cabinetserver), on(videoStorage, cloud)] .
\end{Verbatim}
\vspace{-1mm}
\noindent 
Assuming now that the Cloud datacentre does not offer anymore the software capabilities required by \textsf{\small Video Storage} (i.e. ~\texttt{\small node(cloud, [centos, gcc, make], inf, []).}), we obtain:
\begin{Verbatim}[fontfamily=zi4, fontsize=\footnotesize, frame=single, framesep=1mm, framerule=0.1pt, rulecolor=\color{gray}]
2 ?- fogBrain(vrApp,P).
P = [on(videoStorage, ispdatacentre), on(vrDriver, accesspoint),
     on(sceneSelector, cabinetserver)]
\end{Verbatim}
\vspace{-1mm}
\noindent
This reasoning step suggests migrating the \textsf{\small Video Storage} from the Cloud to the ISP datacentre. 
Such a result was obtained by exploiting the following partially ground query of \texttt{\small placement/7}:
\begin{Verbatim}[fontfamily=zi4, fontsize=\footnotesize, frame=single, framesep=1mm, framerule=0.1pt, rulecolor=\color{gray}]
placement([videoStorage], [(cabinetserver,2),(accesspoint,2)], NewAllocHW, 
    [(accesspoint,cabinetserver,1),(cabinetserver,accesspoint,8)], NewAllocBW,
    [on(vrDriver,accesspoint),on(sceneSelector,cabinetserver)], NewPlacement).
\end{Verbatim}
\noindent obtained as per the \texttt{\small reasoningStep/5} predicate, by removing \textsf{\small Video Storage} from the input partial placement, and from the input hardware and bandwidth allocations, while leaving untouched information related to non-suffering services (viz. \textsf{\small VR Driver} and \textsf{\small Scene Selector}). 
$\hfill \diamond$

\smallskip\noindent
We conclude this section with a note on the time complexity of our continuous reasoning solution. Identifying suffering services and building the partial query for \texttt{\small placement/7} incurs in worst-case $O(S^2)$ time complexity, bounded by the maximum number of service-to-service requirements to check on $S$ services. If all services are eventually migrated, our solution is still worst-case exp-time $O(N^S)$, over an infrastructure with $N$ nodes. More likely, as we will epitomise in Sect.~\ref{sec:experiments}, a new placement will be determined only for a lower number $s$ of services $s < S$, leading to a $O(N^s)<O(N^S)$ time complexity. For instance, when only one service will be migrated, our approach will be worst-case linear-time $O(N)$ and, when only one service-to-service interaction will trigger migration, it will be worst-case quadratic-time $O(N^2)$.%, only migrating the two involved services. 
\vspace{-3mm}
\section{Experimental Results}
\label{sec:experiments}

\noindent In this section, we run \fogbrain against increasing infrastructure sizes to assess the scalability of our continuous reasoning approach to support runtime application management. 

\smallskip\noindent\textit{Dataset. } In all experiments, the application to be placed is the one of Fig. \ref{fig:vrapp}. Target infrastructures replicate the infrastructure of Fig. \ref{fig:vrinfra} for a number $R \in \{2, 10, 20, 100, 200\}$ of times, and fully connect nodes with suitable links\footnote{The Python code which can be used to generate infrastructures of arbitrary sizes by replicating the base module is available at: \url{https://github.com/di-unipi-socc/fogbrain/blob/master/infrastructure_builder/builder.py}}. Finally, infrastructures feature a single smartphone and a single access point capable of reaching out the VR Viewer needed by the application. 

\smallskip
\noindent
\textit{Execution Environment.} We run \fogbrain %\footnote{All \textsf{\scriptsize FogBrain} code available at \url{https://github.com/di-unipi-socc/fogbrain}} 
in SWI-Prolog\footnote{SWI-Prolog available at: \url{https://www.swi-prolog.org/}} 64-bits (v. 8.0.3) on a commodity laptop featuring Windows 10, an Intel i5-6200U CPU (2.30 GHz) and 8 GB of RAM. We rely upon the \texttt{\small time/1} meta-predicate to count the inferences needed by \texttt{\small reasoningStep/5} (line 46) and by \texttt{\small placement/2} (line 49). Counting inferences (instead of only measuring time) permits to assess the performance of continuous reasoning against a machine-independent metric. Finally, we do not consider the inferences needed to load infrastructure data, which we assume can be periodically and incrementally updated by an infrastructure monitoring tool. %(e.g. FogMon \cite{fogmon}).

\smallskip
\noindent
\textit{Experiments. } For all different sizes of the infrastructure we perform the following:
\begin{enumerate}
    \item[(1)] a run of \fogbrain to find a first deployment of the application,
    \item[(2)] a run of \fogbrain where infrastructure changes led to no need for migration,
    \item[(3a)] a run of \fogbrain where the Cloud node exploited for deployment features too low hardware resources (viz. 0) versus a run of \texttt{\small placement/2} over the same infrastructure, and
    \item[(3b)] a run of \fogbrain where, instead, the link between the Cloud node and the cabinet server exploited for deployment in the infrastructure features too high latency (viz. 1350 ms) versus a run of \texttt{\small placement/2} over the same infrastructure.
\end{enumerate}

\noindent\textit{Discussion. } Table \ref{tab:results} shows the results of all experiments (1), (2), (3a) and (3b).
It is worth noting that determining a first placement (1) requires an exponentially increasing number of inferences as the infrastructure size grows. In our settings, this step requires 1 to 2 seconds to find a valid first placement with 2000 infrastructure nodes, which is a tolerable amount of time when the application is not running yet.
However, when it comes to runtime application management, reducing the number of required inferences and, consequently, decision-making times, is crucial to avoid prolonged application performance degradation. 
As per Table \ref{tab:results}, using continuous reasoning to analyse infrastructure changes that do not trigger migrations (2) requires a constant number of inferences, independently of the infrastructure size. In our settings, this corresponds to negligible execution times for all considered infrastructure sizes. This happens because the time complexity of the performed check is at most $O(S^{2})$ for an application with $S$ services (as discussed in Sect. \ref{sec:continuousreasoning}) and $S$ is constant (viz. $3$) throughout the experiments.

Table \ref{tab:results} compares the number of inferences needed to react to (3a) and (3b) with continuous reasoning (i.e. CR Inf.s) to the number of inferences needed to react to them by computing a whole new placement (i.e. No CR Inf.s) using \texttt{\small placement/2}. Table \ref{tab:results} also reports inference speed-ups (viz. (No CR Inf.s $/$ CR Inf.s)) achieved by continuous reasoning for (3a) and (3b).
The speed-up for (3a) is over $5500\times$ with an infrastructure of $1000$ nodes\footnote{The number of CR inferences for (3a) is constant due to the Prolog ordering of the clauses of \texttt{node/4} in the example, which always leads to finding a new placement on a node other than the Cloud node in $369$ inferences. Anyway, as discussed at the end of Sect. \ref{sec:continuousreasoning}, in the worst case, the number of CR inferences increases at most linearly in the number of infrastructure nodes when migrating a single service.}. 
Analogously, in the case of a single link failure (3b), which requires migrating two services and all service-to-service allocations, the speed-up is over $95\times$ with an infrastructure of $1000$ nodes. It is worth noting that in both (3a) and (3b) \fogbrain starts speeding up with at least $10$ nodes since, for smaller infrastructure sizes, analysing the current deployment conditions requires a number of inferences greater than computing a placement anew.
\vspace{-5mm}
 \begin{table}[!ht]
 \centering
 \caption{Experimental results.}\label{tab:results}
\includegraphics[width=0.99\textwidth,trim=13mm 15mm 25mm 12mm, clip]{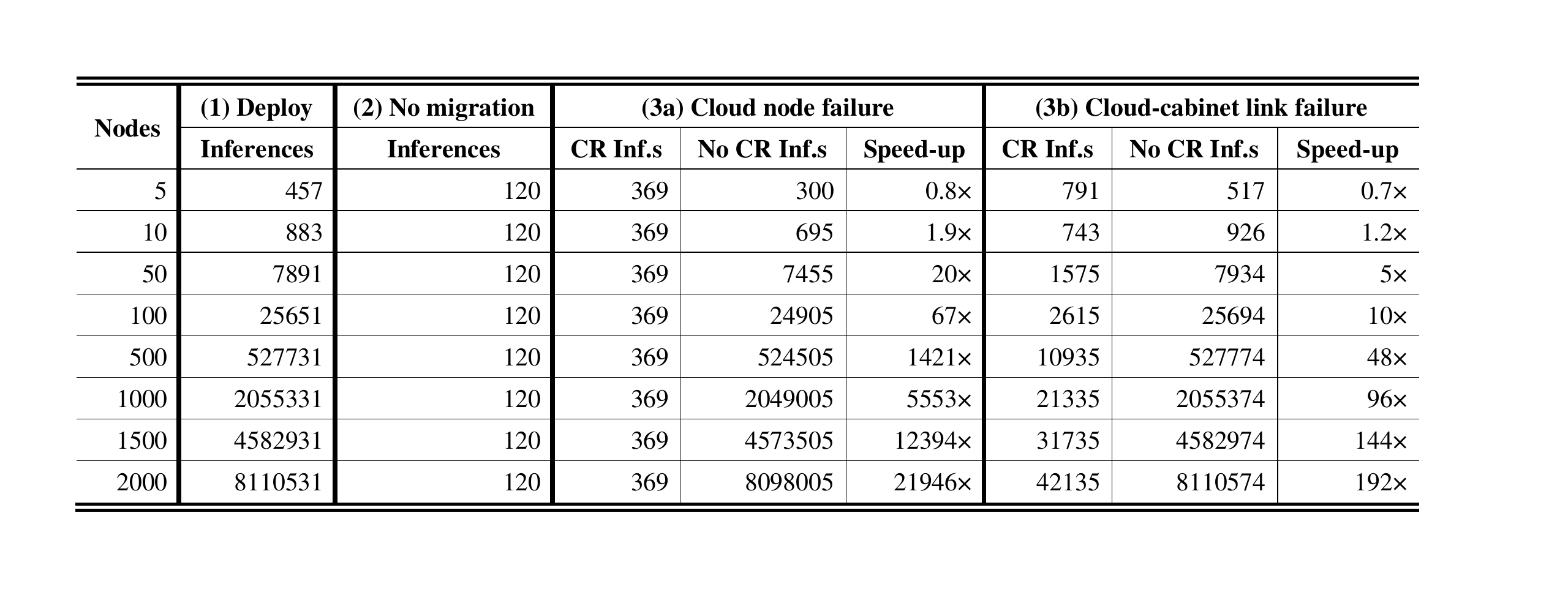}
 \end{table}
\vspace{-3mm}
\newline\noindent Overall, the usage of a continuous reasoning approach to support application placement decision-making shows very promising results in our experiments as it substantially reduces the time needed to make decisions at runtime, in the likely situation where only a portion of the deployed application services cannot currently satisfy its requirements.

\vspace{-3.5mm}
\section{Related Work}
\label{sec:related}
In the past, much literature has focussed on the placement of application services to physical servers in Cloud datacentres~\cite{pietri2016mapping}, only a few of which (e.g.  \cite{kadioglu2016heterogeneous,yin2009rhizoma}) employing a declarative approach.
However, managing applications over the Cloud-IoT continuum introduces new peculiar challenges, mainly due to infrastructure scale and heterogeneity, need for QoS-awareness, dynamicity and support to interactions with the IoT, rarely considered in Cloud-only scenarios. 
Next, we briefly summarise the state of the art in the field of Cloud-IoT multi-service application placement and management, referring the readers to our recent survey~\cite{brogi2019place} for further details. 

Among the first proposals investigating the peculiarities of Cloud-IoT application placement, \cite{004} proposed a simple search algorithm to determine an eligible deployment of (multi-service) applications to tree-like Cloud-IoT infrastructures, open-sourced in the iFogSim Java prototype. Building on top of iFogSim, various works tried to optimise different metrics, e.g. service delivery deadlines~\cite{021}, load-balancing~\cite{032}, or client-server distances~\cite{101}. 
In our previous work \cite{QoSawaredeployment2017,fogtorchpibookchapter2019}, we proved NP-hardness of the placement problem, and we devised a backtracking strategy to determine context-, QoS- and cost-aware placements of multiservice applications to Cloud-IoT infrastructures, also employing genetic algorithms to speed up the search~\cite{geneticalgorithms2019}. Based on our work and on the related FogTorch$\Pi$ Java prototype, \cite{xia2018combining} focussed on minimising application response times, while \cite{demaio} proposed a strategy for Cloud-IoT task offloading. 
Very recently, we exploited logic programming to assess the security and trust levels of application placements \cite{secfog2019}, and to determine the placement and network routing of Virtual Network Function chains in Cloud-IoT scenarios \cite{edgeusher}.
To the best of our knowledge, no other previous work tackling the problem of application placement to Cloud-IoT infrastructures relies on declarative programming solutions. Besides, no previous work proposes continuous reasoning approaches to tame the exp-time worst-case complexity of such a problem, nor to solve it incrementally at application management time.

Finally, proposals exists for simulating application placements and management policies in Cloud-IoT scenarios \cite{margariti2020modeling}, e.g. YAFS \cite{lera2019yafs}, EdgeCloudSim \cite{sonmez2018edgecloudsim}, and iFogSim \cite{004} itself. Recently, we also worked on simulating the management of CISCO FogDirector-enabled infrastructures and opensourced the FogDirSim prototype \cite{fortipagiarobrogi2020}. 
Those simulators were mainly used to assess static placements, or dynamic performances of simple management policies (e.g. random, first-fit, best-fit), reasoning on the whole infrastructure and application status when facing infrastructure changes.

Summing up, to the best of our knowledge, this work is the first proposing a declarative continuous reasoning solution to support placement decisions during runtime management of multi-service applications over Cloud-IoT infrastructures.

\vspace{-3mm}
\section{Concluding Remarks}
\label{sec:conclusions}

In this article, we presented a novel declarative continuous reasoning methodology, and its prototype \fogbrain, to support runtime management decision-making concerning the placement of next-gen multiservice applications to Cloud-IoT infrastructures. 
\fogbrain was assessed over a lifelike example at varying infrastructure sizes, from small-scale to large-scale. Limiting placement decisions only to services affected by the last infrastructure variations via continuous reasoning has shown considerable average speedups (i.e. $\geqslant 2500\times$).
Overall, \fogbrain represents the core of a continuous reasoner to support Cloud-IoT application management which, being declarative, is concise ($125$ single lines of code), and easier to understand and extend so to account for new emerging needs, compared to existing procedural solutions ($\geqslant 1000$s lines of code). 

As future work, we plan to extend the continuous reasoning capabilities of \fogbrain to also handle changes in the application topology (i.e. addition/removal of services) or requirements (e.g. security policies), to support other runtime management decisions (e.g. application scaling, service adaptation), and to exploit a cost-model and heuristic algorithms to determine optimal eligible placements, possibly along with constraint logic programming and incremental tabling. Besides, we intend to include the possibility to obtain textual explanations on \textit{why} a certain management decision was taken by \fogbrain and to exploit probabilistic logic programming to simulate infrastructure variations as per historical monitored data. Last, but not least, we plan to assess \fogbrain over other use cases and in testbed settings over actual multi-service applications.

\vspace{-1mm}
%\section*{Acknowledgements} 
\medskip
\noindent
{\small
{\bf Acknowledgements.} Work partly supported by the projects ``{\it DECLWARE}"(PRA\_2018\_66), funded by the University of Pisa, Italy, and ``{\it GI\`O}% a Fog computing testbed for research \& education}
", funded by the Department of Computer Science, University of Pisa, Italy.
}

%\nocite{*}
\vspace{-5mm}
{\small
\bibliography{biblio}
\bibliographystyle{eptcs}
}

\label{lastpage}
\end{document}